\documentclass[12pt]{article}

\usepackage{scicite}

\usepackage{times}
\usepackage{amsmath}

\usepackage{xspace}
\usepackage{doi}

\renewcommand{\url}[1]{\href{#1}{Link}}
\newenvironment{sciabstract}{%
\begin{quote} \bf}
{\end{quote}}

\usepackage{graphicx}
\usepackage{hyperref}

\usepackage{textcomp}
\usepackage{palatino} % font to avoid the ugly default µ
\usepackage{gensymb}
\usepackage{hyperref}

\usepackage{color}

\newcommand{\LongTermStability}{3$\times$10$^{-10}$~rad.s$^{-1}$}

\newcommand{\ShortTermStability}{$3\times 10^{-8}$~rad.s$^{-1}$.Hz$^{-1/2}$}
\newcommand{\keff}{k_\mathrm{eff}}
\newcommand{\veckeff}{\vec{k}_\mathrm{eff}}
\newcommand{\radps}{rad.s$^{-1}$}

\newcommand{\nradpspsqrtHz}{nrad.s$^{-1}$.Hz$^{-1/2}$}
\newcommand{\ms}{\ \text{ms}}
\newcommand{\SM}{Supplementary Materials} 

\title{
 Interleaved Atom Interferometry for High Sensitivity Inertial Measurements
}

\author
{D. Savoie$^{\dagger}$, M. Altorio$^{\dagger}$, B. Fang, L. A. Sidorenkov, R. Geiger$^{\ast}$, A. Landragin\\
\\
\normalsize{LNE-SYRTE, Observatoire de Paris,  Universit\'e PSL, CNRS, Sorbonne Universit\'e}\\
\normalsize{61 Avenue de l'Observatoire, 75014 Paris, France}\\
\normalsize{$^\dagger$These authors contributed equally to this work.}\\
\normalsize{$^\ast$To whom correspondence should be addressed; E-mail:  remi.geiger@obspm.fr.}
}

\date{}

\begin{document} 


\baselineskip24pt


\maketitle


\begin{sciabstract}
Cold-atom inertial sensors target several applications in  navigation, geoscience and tests of fundamental physics. 
Reaching high sampling rates and high inertial sensitivities, obtained with long interrogation times, represents a challenge for these applications.
 We report on  the  interleaved operation of a cold-atom gyroscope, where 3 atomic clouds are interrogated simultaneously in an atom interferometer featuring a 3.75~Hz sampling rate and an interrogation time of 801~ms.
Interleaving improves the inertial sensitivity by efficiently averaging  vibration noise, and allows us to perform dynamic rotation measurements in a so-far unexplored range.
We demonstrate a stability of \LongTermStability, which competes with the best  stability levels obtained with fiber-optics gyroscopes. 
Our work validates interleaving as a key concept for future atom-interferometry sensors probing time-varying signals, as in on-board navigation and gravity-gradiometry, searches for dark matter, or gravitational wave detection.

\end{sciabstract}

% In setting up this template for *Science* papers, we've used both
% the \section* command and the \paragraph* command for topical
% divisions.  Which you use will of course depend on the type of paper
% you're writing.  Review Articles tend to have displayed headings, for
% which \section* is more appropriate; Research Articles, when they have
% formal topical divisions at all, tend to signal them with bold text
% that runs into the paragraph, for which \paragraph* is the right
% choice.  Either way, use the asterisk (*) modifier, as shown, to

\section*{Introduction}
Quantum sensing  relies on the  manipulation of internal or external degrees of freedom in atoms, molecules, opto-mechanical devices, photonic  or solid-state systems, and covers various applications such as magnetometry \cite{Sheng2013,Gross2017,Jimenez-Martinez2018}, the definition of frequency standards \cite{Takamoto2005,LeTargat2013}, short-range force measurements \cite{Tarallo2014} or electromagnetic measurements \cite{Bagci2014,Facon2016}.
 Inertial sensors based on the coherent manipulation of superpositions of momentum states in atom interferometers  have been developed for more than 25 years \cite{Borde1989,KasevichChu1991,Riehle1991}, with the goal to address various applications. Examples of  remarkable achievements are  tests of fundamental physics  \cite{Bouchendira2011,Lepoutre2012,Zhou2015,Jaffe2017,Asenbaum2017}, metrology \cite{Rosi2014}, or absolute gravimetry \cite{Peters2001,Freier2016,Bidel2018,Karcher2018}. 
Such  precision measurements of gravito-inertial effects directly take  benefit from the inherent accuracy and long-term stability of cold-atom sensors. These two properties can eventually be combined with the high-bandwidth of relative sensors, which is at the basis of sensor fusion \cite{Lautier2014}. 
 This approach is reminiscent of   atomic clocks, where probing the stable atomic energy structure is used for   stabilizing a microwave or optical oscillator \cite{Takamoto2005,LeTargat2013}, or for tests of fundamental physics. 
% To our knowledge, only two demonstrations of a real gain offered by inertial sensors based on atom interferomery have been reported to date:  absolute gravimetry \cite{Karcher2018}, and marine navigation \cite{Bidel2018}.

The extension of  the applications of cold-atom inertial sensors to the measurement of time-varying signals has been confronted to their reduced sampling rate, which originates from their sequential operation and from the long interrogation time of the atoms  that is required to achieve high inertial sensitivity.
This limitation is, for example, an obstacle for applications to inertial navigation \cite{Jekeli2005}  or to fundamental research related to dark matter detection  \cite{Graham2016} or gravitational wave astronomy \cite{Chaibi2016,Graham2016a}.
% A major obstacle for applications  has been  the reduced sampling rate of  cold-atom  sensors associated with dead times, which results in a degraded sensitivity due to vibration noise aliasing \cite{Jekeli2005}. 
% Besides  inertial navigation, achieving higher sampling rate and high sensitivity is a key feature for applications where signals varying on second time-scales or faster are of interest, as, for example, in on-board gravity-gradiometers. 
In this work, we report on the  interleaved operation of a cold-atom inertial sensor, which operates with a sampling frequency of $3.75$~Hz and features a high inertial sensitivity, as given by the 801~ms interrogation time of the atoms  in the interferometer.
The method of interleaving, which we demonstrate  for both static and  dynamic rotation rate measurements, can be generalized to other atom interferometer architectures,  and therefore paves the way to the development of high-bandwidth and high-sensitivity cold-atom inertial sensors.

 Besides an increase in sensor bandwidth, we show that interleaving allows  to efficiently average  vibration noise (as $1/\tau$, with $\tau$  the integration time), which represents the most important noise source in cold-atom inertial sensors. 
As a consequence, we demonstrate a   record rotation rate sensitivity of \ShortTermStability. 
Such a high sensitivity level  allows us to  characterize the systematic effects of a cold-atom gyroscope in a so-far unexplored range \cite{Gauguet2009,Berg2015}, and to stabilize them at  the few  $10^{-10}$~rad.s$^{-1}$ level.
Previous works on atomic beam gyroscopes already demonstrated excellent sensitivities \cite{Gustavson2000} and long term stabilities close to the state-of-the-art of optical gyroscopes \cite{Durfee2006}.
As the long-term instability of gyroscopes  is a limiting factor in inertial navigation systems, achieving the performance of  the best fiber-optics gyroscopes \cite{Lefevre2014} was a long-standing goal, which we attain for the first time with a cold-atom sensor.

\section*{Results}

\subsection*{Experimental setup}

\begin{figure}[!h]
\includegraphics[width=\linewidth]{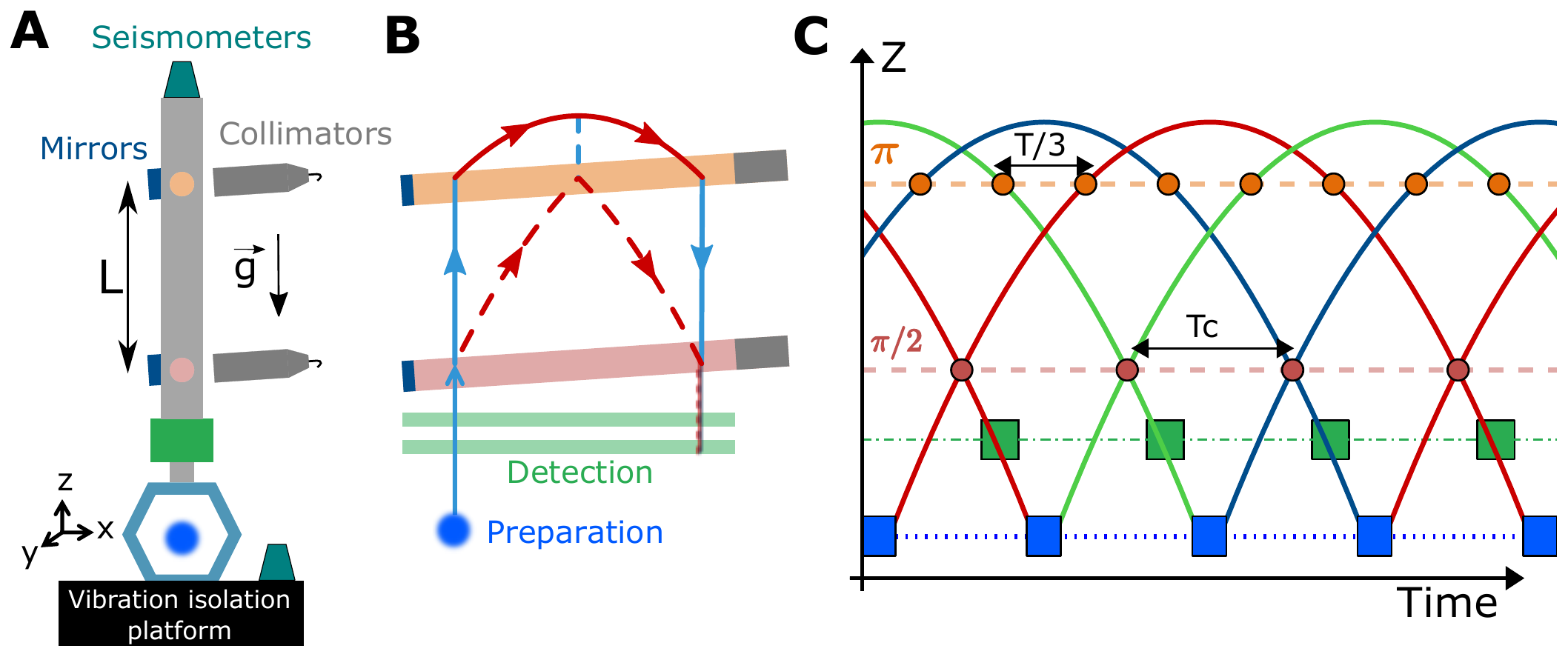}
\caption{\label{fig:detailSequence3IJ}  \textbf{Principle of the experiment}. 
\textbf{A}: sketch of the experiment, where the atoms are laser cooled (blue cloud) and launched vertically,  interrogated by two  Raman beams (brought from the gray collimators and retro-reflected on the blue mirrors), and detected on their way down (green box). The distance between the Raman beams is $L=\frac{3}{8}gT^2\simeq 59$~cm.
\textbf{B}:  diagram of the atom interferometer in the $(xz)$ plane (not to scale), with the blue and red lines respectively labeling  the   $|\vec{0}\rangle$ and $\hbar\veckeff$ momentum states. The dashed and plain lines show the two paths of the matter-waves in the interferometer, which  enclose an area of 11~cm$^2$. 
\textbf{C}: trajectories of the successively launched  atom clouds in interleaved operation.
Each interferometer has an interrogation time  $2T=801$~ms and  the cycle time is $T_c = 2T/3=267$~ms. The  $\pi/2$ pulses are shared between the atom clouds entering and exiting the interferometer. }
\end{figure}

\textit{Experimental sequence and principle of the gyroscope.}
The core of the experimental setup used in this work has been described in \cite{Dutta2016}, and is sketched in Fig.~\ref{fig:detailSequence3IJ}. The essential techniques are given in the Materials and Methods section, with further details in the \SM. In short, we laser-cool Cesium atoms to a temperature of $1.2 \ \mu$K and launch them vertically at a velocity of $5.0$~m.s$^{-1}$. After a  selection  step of the $m_F=0$ magnetic sublevel, we interrogate the atoms in the interferometer, and finally detect their state at the output of the interferometer, on their way down, using fluorescence detection.
We realize the light-pulse atom interferometer using two-photon stimulated Raman transitions with counter-propagating laser beams which couple the $|F=3,m_F=0\rangle$ and $|F=4,m_F=0\rangle$ clock states of the Cesium atom.

According to the Sagnac effect, the rotation sensitivity  is proportional to the area between the 2 arms of the interferometer. 
Our  gyroscope is based on a fountain configuration with 4 light pulses to create a folded geometry thanks to gravity \cite{Canuel2006}.
The symmetric 4 pulse fountain configuration allows to achieve a large  area ($11$~cm$^2$ in this work) and  leads to a vanishing  sensitivity to constant linear accelerations. 
The interferometer phase shift, $\Phi$, can be calculated from the relative phase  between the two Raman lasers, $\Delta\varphi_{\text{laser}}(t)=\vec{k}_{\mathrm{eff}}\cdot \vec{r}_{b,t}(t) + \Delta\varphi(t)$, which is imprinted on the diffracted part of the matter-wave at the time $t$ of the pulse. It reads:
\begin{equation}
\label{eqn:rotphase4p}
\Phi= \vec{k}_{\mathrm{eff}}\cdot \left[ \vec{r}_b \left(0\right) - 2\vec{r}_t \left(\frac{T}{2}\right) + 2\vec{r}_t \left( \frac{3T}{2}\right) - \vec{r}_b \left( 2T\right) \right] + \Delta\Phi^0,
\end{equation}
where $\vec{k}_{\mathrm{eff}}$ is the two-photon wavevector, $\vec{r}_{b,t}(t)$ is the position of the  mirror retro-reflecting the Raman lasers with respect to the center-of-mass of the free falling atoms (subscripts $\{b,t\}$ for bottom and top mirror, see Fig.~\ref{fig:detailSequence3IJ}), and $2T$ is the total interrogation time. 
The last term $\Delta\Phi^0$ is a controllable laser phase shift independent of inertial effects.
The phase shift associated to the stationary Earth rotation rate $\vec{\Omega}_E$ is given by 
\begin{equation}\
\Phi_\Omega=  \frac{1}{2}\vec{k}_{\mathrm{eff}}\cdot \left(\vec{g}\times\vec{\Omega}_E\right) T^3,
\label{eq:constant_rot}
\end{equation}
where $\vec{g}$ is the acceleration of gravity \cite{Stockton2011}.

\textit{Interleaved operation.}
We employ a sequence of joint interrogation of successive interferometers, which is obtained by  using the same $\pi/2$ Raman pulse for the atom clouds entering  and exiting the interferometer zone \cite{Dutta2016}. 
Consequently, the sensor can operate without dead times.
The interleaved operation, which is reminiscent from the atom juggling technique of Ref.~\cite{Legere1998}, is then implemented by extending this joint sequence to a multiple-joint sequence, as proposed in  \cite{Meunier2014}.
The sequence of Raman pulses is given in Fig.~\ref{fig:detailSequence3IJ}.
 If we denote $2T=801$~ms the total duration of the interferometer, we launch an atom cloud every $T_c=2T/3=267$~ms, which supposes that a cloud is laser-cooled while 3  previously launched  clouds are interrogated in the interferometer.
Due to  timing constraints, the loading time of the magneto-optical trap (MOT) is limited. 
The atoms are loaded in the MOT during  55~ms, and we  detect $2\times$10$^5$ atoms at the end of the interferometer. 
The light scattered from the MOT atoms causes incoherent photon absorption and emission from the interrogated atoms and therefore a loss of contrast  \cite{Meunier2014}. 
The  contrast of the interferometer is 7.4$\%$, limited by the expansion of the cloud during the free fall  in the   Raman beams of gaussian profile, and by the  light scattered from the MOT.  

\textit{Technical upgrades.}
We implemented  several key upgrades of our setup compared to Ref.~\cite{Dutta2016}. First, we improved the detection noise   which was limiting the sensitivity in \cite{Dutta2016}. The equivalent one-shot phase noise is now 71~mrad, corresponding to a rotation noise of 8 \nradpspsqrtHz.
 Second, we implemented a real-time compensation of linear acceleration noise  \cite{Lautier2014}, and a servo-loop to operate the interferometer at mid-fringe, i.e. in its linear range. These techniques are described in  the Materials and Methods section.
 These upgrades result in a sensor which  effectively operates without dead times, as statistically very few points sit on the top or bottom of a fringe, where the sensitivity vanishes.

\begin{figure}[!h]
\includegraphics[width=\linewidth]{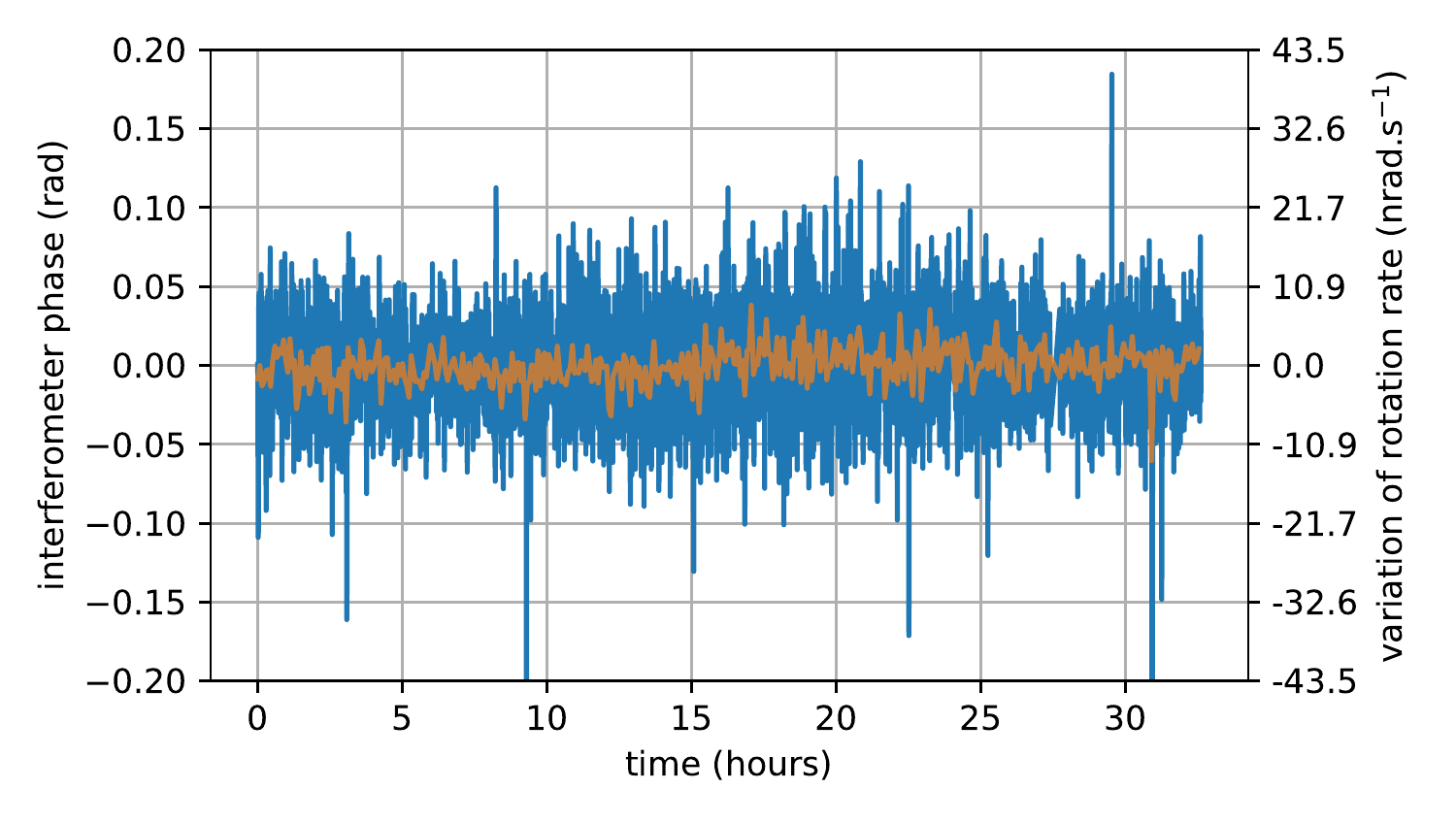}
\caption{\label{fig:long_acq}\textbf{ Rotation rate  measurement during 32.5 hours. }
% data set 13 =  21:10 on 23/09 to 5:41 on 25/09
In the blue (orange) trace, each data point is the average over segments of 26.7~s (267~s)  of  raw  inertial phase measurements.
The right axis translates inertial phase to rotation rate using the scale factor of the gyroscope to stationary Earth rotation (from Eq.~\eqref{eq:constant_rot}).
% The plain line shows a sinusoidal fit to the data  with fixed 24 hour period and an extracted amplitude of 4.1~mrad.
 }
\end{figure}

\textit{Rotation rate acquisition.}
Fig.~\ref{fig:long_acq} shows a 32.5 hours acquisition of rotation rate measurements obtained between September 23rd and 25th, 2017. To obtain this series of data, we alternated the direction of the Raman wavevector  ($\pm \veckeff$) and computed the half-difference of two successive measurements to reject non-inertial ($\veckeff$-independent) effects, such as AC Stark shifts (see Methods and \SM - section S1 for the details of the sequence, and  section S2 for the raw data).
In the following, we will analyze the sensitivity and the stability of the gyroscope from this acquisition.

\subsection*{Efficient  averaging of vibration noise and record sensitivity}

Vibration noise is the most important source of sensitivity degradation in cold-atom inertial sensors of large area (i.e. using long interrogation time and/or large momentum transfer techniques \cite{Dickerson2013}).   
Efficient vibration isolation at low frequencies (below few Hz) is technically challenging (e.g. \cite{Hensley1999}) and not suited for field applications. We will show that interleaving allows to reduce the impact of this key noise source.

In our sensor, the impact of inertial noise can be analyzed by considering a center of rotation located at the top Raman beam: inertial noise then appears as linear acceleration noise of both mirrors, plus rotation noise of the bottom mirror. 
The rotation noise translates into random variations of the angle $\theta_B(t)$ of the Raman beam with respect to a geostationary reference frame \cite{Stockton2011},  and impacts the interferometer phase  as $\left [\theta_B(2T)-\theta_B(0)\right]$ (Eq.~\eqref{eqn:rotphase4p}).
In joint measurements, in which $\pi/2$ pulses are shared (occurring at times $0$ and $2T$), the contribution of rotation noise  cancels out when  averaging $N$ successive measurements (see Material and Methods for a derivation).
Therefore,  the gyroscope sensitivity should improve as $\tau^{-1}$, where $\tau=2 N T$ is the integration time, instead of $\tau^{-1/2}$ in the case of uncorrelated measurements impacted by  rotation noise.

Besides averaging rotation noise, the interleaved operation of our sensor allows us to reduce the impact of residual linear acceleration noise: because our sampling frequency ($1/T_c=3.75$~Hz) is higher than the  frequencies at which  the acceleration noise mostly contributes (around 0.5~Hz, see Table S1 in \SM), correlations  appear between successive measurements, yielding a scaling of the sensitivity that approaches $\tau^{-1}$ (rather than $\tau^{-1/2}$).

%which pushes towards a $\tau^{-1}$ (rather than $\tau^{-1/2}$)} dependence of the  sensitivity

\begin{figure}[!h]
\includegraphics[width=\linewidth]{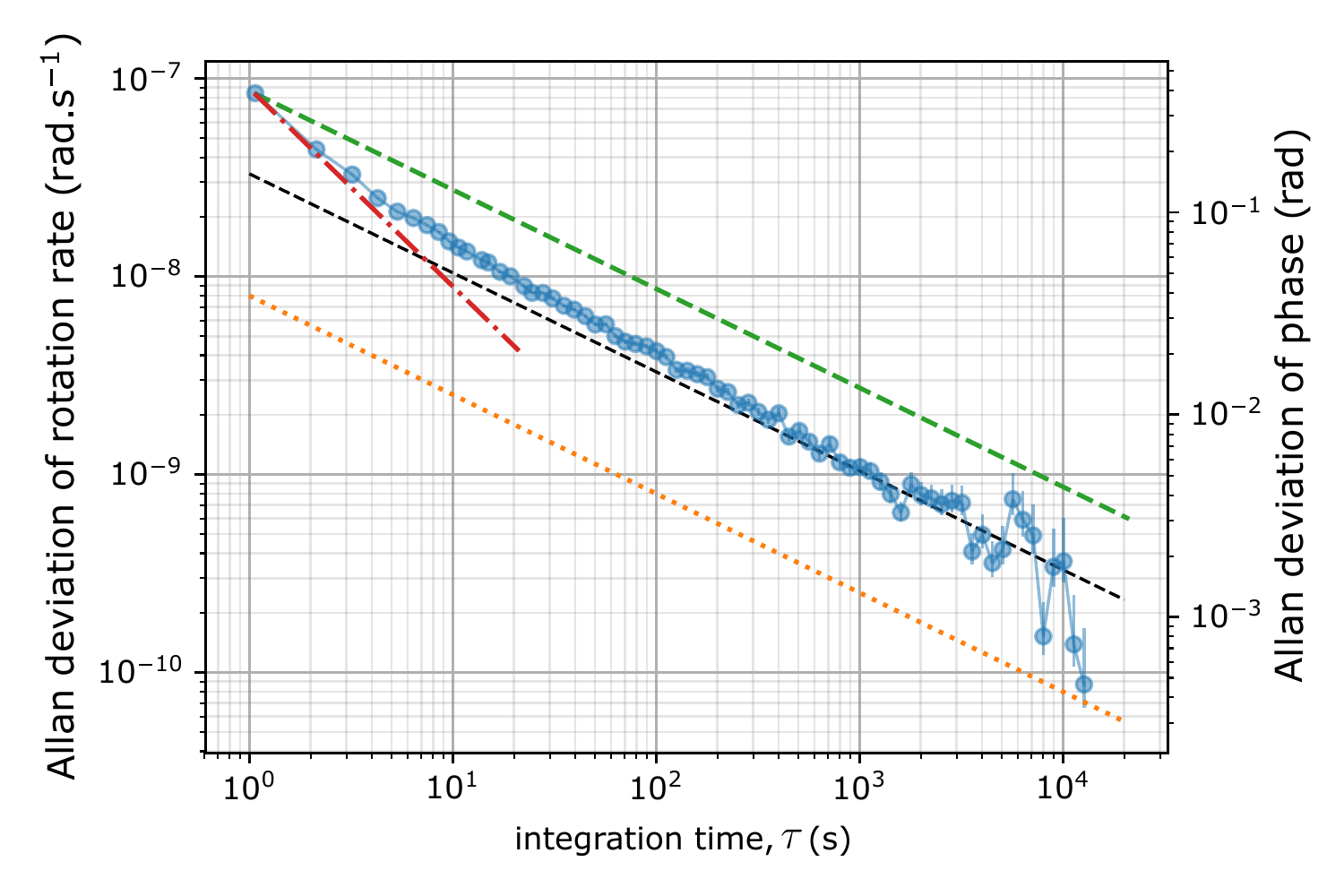}
% 12: 1:22 - 12:47 on 24/09/17
\caption{\label{fig:short_term} \textbf{Gyroscope sensitivity.} Stability analysis of a 11.3 hour portion of the rotation rate measurements of Fig.~\ref{fig:long_acq}, between 1:22 and 12:47 on September 24th, 2017. The error bars represent the 68~$\%$ confidence intervals on the estimation of the Allan deviation.
Dashed black line: $3.3\times 10^{-8}$~rad.s$^{-1}\times \tau^{-1/2}$. 
Green dashed  line:  $\tau^{-1/2}$ scaling from the one shot Allan deviation. 
Red dotted-dashed  line:  $\tau^{-1}$ scaling from the one shot Allan deviation. 
Orange dotted line: detection noise limit corresponding to $8\times 10^{-9}$~rad.s$^{-1}\times \tau^{-1/2}$.
}
\end{figure}

Fig.~\ref{fig:short_term} shows the Allan deviation of the gyroscope stability for a  11.3 hour portion of night data of Fig.~\ref{fig:long_acq}. The improvement of the sensitivity as $\tau^{-1}$ for  integration times up to $\simeq 7$~s is clear. The stability then gradually enters the  $\tau^{-1/2}$ regime characteristic of uncorrelated white noise, corresponding to a sensitivity of \ShortTermStability.
This sensitivity, which improves by more than a factor of 3 on our previous result \cite{Dutta2016}, establishes the  new record for cold atom gyroscopes. 
%In other words, the sensitivity level of 1~\nradps \ is now reached 10 times faster than  in our previous experiment.
As a comparison, our short term sensitivity competes favorably with that of the best fiber-optics gyroscopes \cite{Lefevre2014}.
Such sensitivity enables to study several systematic effects affecting a cold atom gyroscope, for the first time in the range of low $10^{-9}$~rad.s$^{-1}$.

\subsection*{Systematic effects and gyroscope long term stability}

A  systematic shift specific to the interleaved interrogation originates from the  light scattered from the MOT towards the atoms interrogated in the interferometer \cite{Meunier2014}. 
 The MOT scattered light is close to resonance  and induces a loss of contrast and a differential light shift (AC Stark shift). 
The influence of induced light shifts  is  reduced by the spin-echo-like four-pulse sequence, and   by the use of $\veckeff$ reversal: alternating  $\pm\hbar\veckeff$ momentum transfers changes the sign of the inertial phase shift  but not the one of the clock-terms (e.g. differential light shift), which are rejected when taking the half-difference of two measurements (as done in Fig.~\ref{fig:long_acq}).
We measured the residual effect, and showed that it corresponds to an instability below  $7\times 10^{-11}$~\radps \ (see \SM). 
Although currently negligible, this effect is purely technical, and could be resolved by having the MOT and the detection region out  of view from the atom interferometer region in future designs.

%The detailed study of the other systematic effects will be reported in a dedicated paper. 
The most important systematic effects in atom interferometers with separated Raman beams originate from relative wavefront mismatch coupled to deviations of the atom trajectories with respect to the ideal one \cite{Gauguet2009,Tackmann2012}.
In our system, a relative angular misalignement $\vec{\delta\theta}$ between the top and bottom mirrors  used to retro-reflect the Raman beams (Fig.~\ref{fig:detailSequence3IJ}), coupled with an error of launch velocity $\vec{\delta v}$ (with respect to a velocity of $-\vec{g} \ T$ at the first Raman pulse) in the $(y,z)$ plane results in a phase shift 
\begin{equation}
\Delta\Phi = 2T\keff \left(\delta v_y \delta \theta_y + \delta v_z \delta \theta_z \right) = 12 \ \text{mrad}\times \left (\frac{\delta v_{y,z}}{1 \ \text{mm.s}^{-1}}\right) \times \left(\frac{\delta\theta_{y,z}}{1 \ \mu\text{rad}}\right ).
\label{eq:shift_relative_tilt}
\end{equation}
%where 12~mrad of phase shift represents a rotation rate of $2.6$~nrad.s$^{-1}$ (from Eq.~\eqref{eq:constant_rot}).
We explain in the Materials and Methods section how we set the parallelism between the two Raman beams and the velocity of the atoms to approach the ideal trajectory, in order to achieve an uncertainty on the residual systematic shift of  21~mrad (i.e. $4.6$~nrad.s$^{-1}$, from Eq.~\eqref{eq:constant_rot})).

%\textit{Stability analysis.}
After this systematic analysis and the corresponding fine-tuning of the apparatus, we recorded the rotation rate acquisition displayed on Fig.~\ref{fig:long_acq}. 
The  stability of the gyroscope over the entire acquisition is  analyzed in the \SM \ (Fig.~S5), and is in agreement with that read from Fig.~\ref{fig:short_term} for shorter integration times.

\subsection*{Dynamic rotation rate measurements.}

\begin{figure}[!h]
\includegraphics[width=\linewidth]{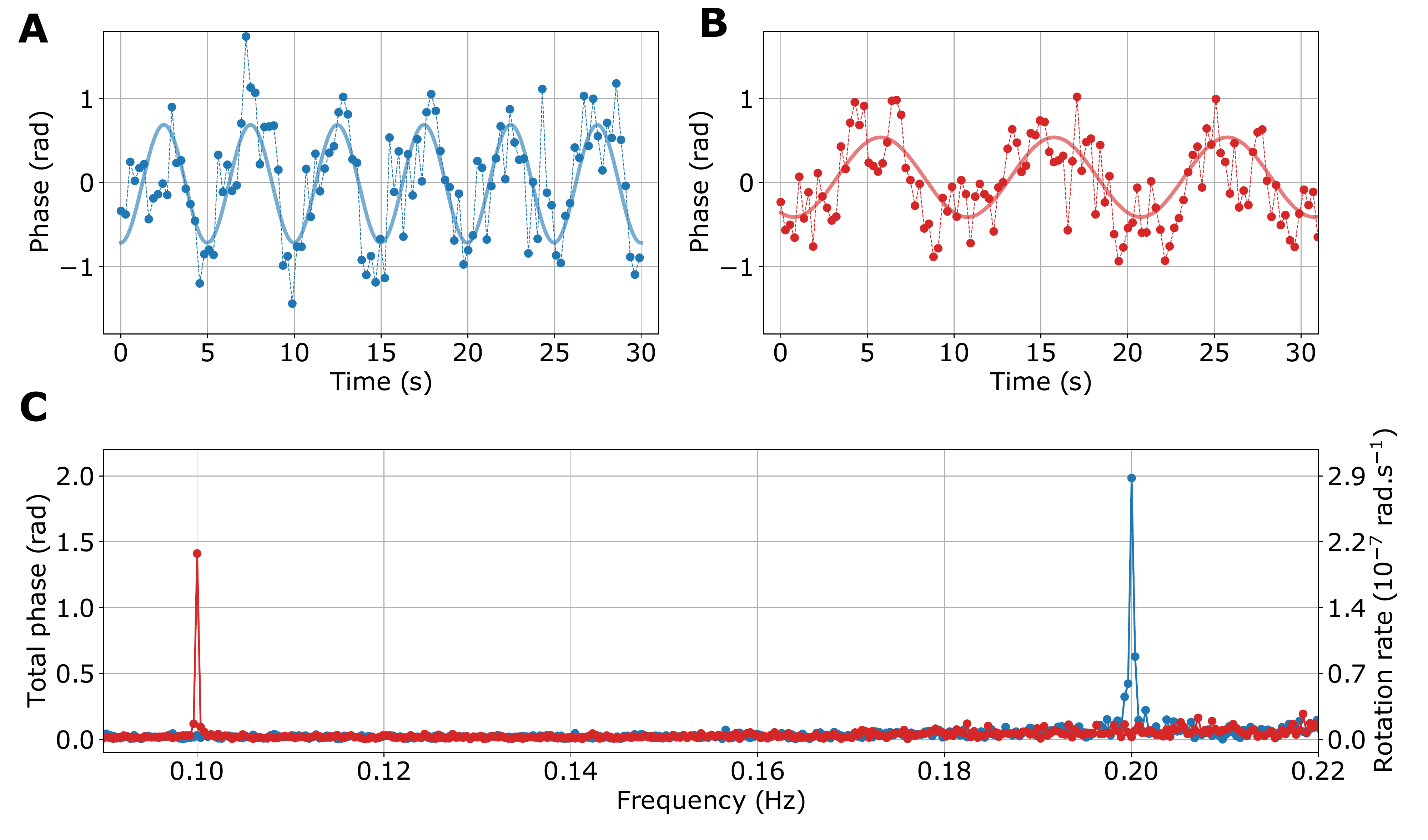}
% Alltogheter_Fit_FFT_5_10_15_second.ipynb
% http://localhost:8888/notebooks/Data_analysis/2018/progs/Matteo/SavedAnalysis/Modulation/Alltogheter_Fit_FFT_5_10_15_second-remi-v3.ipynb
\caption{\label{fig:modulation} \textbf{Measurement of dynamic rotation rates.}  Atom interferometer phase deduced from the transition probability, for rotation rate modulations of 5~s period (A) and 10~s  period (B). Plain line: sinusoidal fit to guide the eye.
C): Fourier analysis of the total rotation rate signal, with a frequency resolution is 0.37~mHz.
}
\end{figure}

We use the unprecedented  sampling rate and  inertial sensitivity of our gyroscope to perform measurements of  weak dynamic  rotation rates. To this end, we  modulate the orientation of the experiment around the $y$ axis. This was performed by applying a force on the bottom plate linking the experimental frame to the vibration isolation platform, via the voice-coil actuator controlling the tilt $\theta_x$ of the apparatus. We apply sinusoidal modulations of the form $\theta_x(t)=\theta_{0}\sin(\omega t)$ with a period  $2\pi/\omega$ and with an amplitude $\theta_{0}$ of few $10^{-7}$~rad.
  The resulting  rotation rate is of the form $\vec{\Omega}(t)=\Omega_0\cos(\omega t) \hat{u}_y$, with $\Omega_0=\omega\theta_{0}$.
The measurements are reported in Fig.~\ref{fig:modulation} for modulation periods of 5~s  and 10~s. The respective modulation amplitudes are $2.3\times 10^{-7}$~rad and $3.4\times 10^{-7}$~rad.
Panels A and B show the atomic phase extracted from the transition probability, $P(t)$, which follows the sinusoidal modulation. The total rotation signal from the atom interferometer is the sum of this atomic phase and the phase compensated in real time. A Fourier analysis of the total signal is shown in panel C. 
Within our frequency resolution, we find that the amplitude of the reconstructed rotation rate signal agrees with the expectation of $\Omega_0$ with a relative precision of $5\%$.
A more detailed  analysis is presented in the \SM \ (section~S5). 
Our proof-of-principle experiment, performed in a so-far unexplored range of time resolution and inertial sensitivity for a cold-atom sensor, demonstrates the impact of interleaved atom interferometry for dynamic measurements.

\section*{Discussion}
\label{discussion}

We demonstrated the method of interleaving in a large-area atom interferometer, as a way to reach high sampling frequencies and high inertial sensitivities together.  
Interleaving enables to efficiently average  vibration noise (the largest noise source in cold-atom inertial sensors), and is thus a promising way of reaching the quantum projection noise limit, a necessary condition before increasing the atom flux or implementing schemes to approach the Heisenberg limit. 
As a result, we  demonstrated record short term sensitivities for a cold-atom gyroscope, and could thus characterize systematic effects in a so-far unexplored range. The rotation rate sensitivity and stability which we achieved  competes  with that of the best strategic-grade fiber-optics gyroscopes (long-term stability in the range of $5 \times 10^{-10}$~rad.s$^{-1}$ \cite{Lefevre2014}). 
Our results thus  pave the way for a change of technology in future high-precision inertial navigation systems.

In our setup, the maximum number of interleaved measurements is technically limited to 3  because of the arrangement of our detection system with respect to the MOT region (see Materials and Methods). In a dedicated design, e.g. where the detection region would be out of view from the upcoming clouds, sampling frequencies of 20~Hz or higher could be reached. 
As an alternative, the use of atoms characterized by different transition wavelengths for the cooling/detection/atom interferometer would be  beneficial, in order to circumvent the effects associated with the scattered light from the source or the detected atoms. Our technique is thus well-suited to ongoing developments of atom interferometers with alkaline-Earth atoms \cite{Hu2017}.

Interleaving ties well with laser-cooling techniques, which are able to rapidly (in less than $100$~ms) produce  cold samples with more than $10^7$ atoms. 
Laser cooling beyond optical molasses such as degenerate Raman sideband cooling appear as a suitable solution for an increased brightness without compromising the cycling frequency. Interleaving is, in principle, also  compatible with the production of ultracold, collimated, atom sources \cite{Asenbaum2017}, provided they can be produced  \cite{Rudolph2015} or extracted at sufficiently high (several Hz) repetition rates.

The method of interleaved atom interferometry can be applied to different sensor architectures, such as multi-axes accelerometers (by alternating measurements along different axes at a high repetition rate), gavimeters or gradiometers. 
For example, interleaving can be exploited to realize a gravimeter of both high accuracy and high sensitivity in a single instrument, potentially allowing to surpass  superconducting gravimeters that currently feature record sensitivities but require regular calibrations. 
As such, interleaving is representative of the flexibility of cold atoms for realizing versatile inertial sensors, as compared to architectures involving macroscopic masses and  electro-mechanical systems.
Regarding fundamental physics applications, reaching high sampling rates is a prerequisite for future studies on dark-matter with atomic accelerometers \cite{Graham2016}, as well as for gravitational wave detection with atom interferometers \cite{Chaibi2016,Graham2016a}.
Interleaving is  therefore  a key concept for future applications of cold-atom inertial sensors.

% **************************************

\section*{Materials and Methods}

\subsection*{Details of the experiment}

%\textit{Experimental sequence.}
Cesium atoms loaded from a 2D Magneto-Optical Trap (MOT) are trapped and laser-cooled in a 3D-MOT. We launch the atoms   vertically at  a velocity of $5.0 \ \text{m.s}^{-1}$ using moving molasses  with a (3D) cloud  temperature of  $1.2 \ \mu\text{K}$.  
After the MOT and prior to the interrogation, the atoms are prepared in the $|F=4,m_F=0\rangle$ state using a selection scheme based on the Stern-Gerlach effect (magnetic deflection of the atoms in $m_F\neq 0$ states).
Light pulse interferometry is realized using two phase-locked Raman lasers  which couple the Cesium clock  states (hyperfine splitting of  $9.192 \ \text{GHz}$). 
The Raman lasers have a wavelength close to the  $D_2$ line (wavelength $\lambda \simeq 852$~nm) and are detuned by $470$~MHz from the excited state to reduce incoherent scattering.
The impact of residual relative Raman laser phase noise has been estimated to 50~mrad per shot of atom interferometer phase.
The Raman lasers are sent to the atoms through two optical windows separated by $L=\frac{3}{8}gT^2\simeq 59$~cm, with an interrogation time  $2T=801 \ms$. 
 We use gaussian Raman beams with $1/e^2$ diameter equal to $40 \ \text{mm}$ and  about 120 mW of total power.
 The interferometer output signal is determined by the probability of transition, $P$, from the $F=4$ to the $F=3$ state, which is read out via fluorescence detection of the two levels' populations after the atom interferometer light-pulse sequence. The probability of transition is modulated according to $P=P_0+A\sin\Phi$, where $C=2A$ is the interferometer contrast and $\Phi$ the interferometer phase.

Our experiment uses retro-reflected Raman beams, such as to form two pairs of Raman beams inducing two transitions: one in the $+\veckeff$ direction, and another  in the $-\veckeff$ direction.
Selectivity of the $\pm \veckeff$ transitions is provided by tilting the Raman beams by an angle $\theta\simeq 3.80$ degrees with respect to the horizontal, in order to introduce a Doppler shift ($\pm \keff g T \sin\theta /2\pi \simeq \pm 611$~kHz at the first and last $\pi/2$ pulses) which is much larger than the  width of the atom Doppler distribution ($\sim 40$~kHz). 
To follow the resonance condition at each Raman pulse, we step-wise change the relative frequency between the two Raman lasers during the sequence, such as to match the values given by the underlying frequency chirp pattern (see details in Fig.~S2 of \SM). To apply the frequency steps, we use a direct digital synthesizer (DDS) driven by an FPGA.

\textit{Real-time compensation of vibration noise and  mid-fringe lock.} We measure the vibrations of the setup with two  broadband seismometers (model \textit{Trillium Compact 120~s }from \textit{Nanometrics}) located at the bottom and top of the experimental frame (see Fig.~\ref{fig:detailSequence3IJ}). From the measured signal, we estimate the interferometer phase shift due to vibrations and apply a corresponding phase jump to the relative phase of the Raman lasers 15~ms before the last pulse.
 This allows us to reduce the standard deviation of the interferometer phase from about 3.2~rad to 0.5~rad. 
To work within the linear regime where the sensitivity is maximal,  we alternate measurements on both sides of a fringe, and compute an error signal from two successive measurements of the  transition probability. This error signal is integrated and used  to servo-lock the interferometer at mid-fringe, via a feedback on the Raman laser relative phase.
More details are given in the \SM, section S1.

\subsection*{Efficient averaging of vibration noise}
Following Eq.~\eqref{eqn:rotphase4p}, and assuming that the Raman lasers are oriented purely in the $x$ direction, the 4-light-pulse atom interferometer phase shift is given by (we neglect the duration of the Raman pulse): 
\begin{equation}
\Phi  =  \keff \left[ x_b(0)-2x_t(T/2)+2x_t(3T/2)-x_b(2T)\right],  
\end{equation}
with $x_{b,t}(t)$ the position of the bottom and top retro-mirrors with respect to the free-falling atom cloud.
The phase shift can be re-written as
\begin{eqnarray}
 \Phi & = &\keff\left[x_t(0)-2x_t(T/2)+2x_t(3T/2)-x_t(2T)\right] + \keff \Big( [x_b(0)-x_t(0)] - [x_b(2T)-x_t(2T)] \Big) \nonumber \\
 & = & \Phi^{\text{acc}}_t + \keff L \left( \theta_b(0)-\theta_b(2T) \right ),
\label{eq:acc_rot}
\end{eqnarray}
with $L=\frac{3}{8}gT^2$ the distance between the bottom and top mirrors, and $\Phi^{\text{acc}}_t$ the term associated to the linear acceleration of the top mirror.
The second term  represents pure rotation of the bottom mirror about the position of the top one.
%If we  consider a pure rotation, with a  center of rotation located at the position of the top mirror, we can neglect the linear acceleration term. 
Recalling that $T_c=2T/3$ and writting as $\Phi_i=\Phi(iT_c)$ the atom interferometer phase at cycle $i$, the mean phase after $N$ measurement   reads
\begin{equation}
\bar{\Phi}_N  = \frac{1}{N}\sum_{i=0} ^{N-1} \Phi_i = \frac{1}{N} \sum_{i=0} ^{N-1} \left( k_{\text{eff}}L \left[\theta_b(iT_c)-\theta_b((i+3)T_c) \right] + \delta\tilde{\phi}_i \right ).
\label{eq:mean_phase}
\end{equation}
The term $\delta\tilde{\phi}_i$ encompasses contributions of detection noise, uncompensated linear acceleration noise and laser phase noise.
When expanding the sum in Eq.~\eqref{eq:mean_phase}, most of the $\theta_b$ terms mutually cancel, such that the mean phase reads
\begin{equation}
\bar{\Phi}_N=k_{\text{eff}}L\frac{\theta_b\left(0\right)-\theta_b\left( (N+2)T_c\right)}{N} + \frac{1}{N} \sum_{i=0} ^{N-1}  \delta\tilde{\phi}_i.
\label{eq:avg_phase}
\end{equation}
This equation shows that the random rotation noise  averages as $N^{-1}$ (first term). The second term represents the uncorrelated noise contributions of  standard deviation $\sigma_{\delta\phi}$. Their sum equals $\sqrt{N}\times\sigma_{\delta\phi}$, which corresponds to a scaling of the phase sensitivity as  as $N^{-1/2}$.

Besides rotation noise,  uncompensated linear accelerations in the frequency range $[0.1 - 1]$~Hz  contribute to a large part the interferometer phase noise (see the \SM \ section S3 for details). This contribution, estimated to typically about 500~mrad per shot, dominates the noise budget and may prevent from  observing a clear $\tau^{-1}$ scaling of the gyroscope sensitivity. 
%The contribution from linear accelerations which are not compensated  is estimated to be  the dominant noise source, representing typically about 500~mrad of phase noise per shot.
%Besides rotation noise, uncompensated linear acceleration noise stems from the frequency band $\sim [0.1 - 1]$~Hz, which is not well captured by seismometers. This contribution, estimated to typically about 500mrad per shot (see SM), dominates the noise and may prevent observing a clear tau-1 scaling of the sensitivity. 
Interleaving, however, allows to over-sample these fluctuations, thus  introducing   correlations between successive measurements, which also contributes to the $\tau^{-1}$ dependence of the instrument sensitivity.

 \subsection*{Alignment of the two Raman beams and atom trajectory}
We set the parallelism between the top and bottom Raman beams by means of a  two-axis piezo-motorized mirror mount with $0.7 \ \mu$rad resolution.  
By optimizing the contrast of the interferometer, we approach the parallelism with an uncertainty of about $3 \ \mu$rad, which is required for the matter-waves to recombine at the output of the interferometer.
For the fine adjustment, we measure the dependence of the phase shift of Eq.~\eqref{eq:shift_relative_tilt}, $\Delta\Phi=2Tk_{\text{eff}} (\delta v_y\delta\theta_y+\delta v_z\delta \theta_z)$, on $\delta \theta_{y,z}$ and $\delta v_{y,z}$ (as defined in the main text).
To this end, we set the atom trajectory in the $(y,z)$ directions by varying the tilt of the experiment ($y$ direction) and the launch velocity during the moving molasses phase ($z$ direction).
In the $z$ direction, we could zero the systematic effect with an uncertainty of $5$~mrad. This amounts to set the velocity of the atoms at the first Raman pulse  to the ideal velocity ($v_z=gT$) with an uncertainty of $0.6$~mm.s$^{-1}$, and to set   the parallelism between two mirrors in the $z$ direction with an uncertainty of $0.7 \ \mu$rad.

The minimization of the systematic shift in the $y$ direction was technically more difficult to achieve than in the $z$ direction: recording the dependence of the phase shift on $\delta\theta_y$ for various velocities required to tilt the entire apparatus by several mrad in order to vary $\delta v_y$ by several mm.s$^{-1}$. This procedure required to manually move masses on the base plate of the experiment sitting on a floating vibration isolation platform, which introduced instabilities. 
We managed to set the $y$-velocity close to the ideal  velocity  ($v_y=0$) with an uncertainty  of 1.8~mm.s$^{-1}$.
The residual shift corresponds to a phase variation of $21$~mrad per $\mu$rad of $\delta \theta_y$ variation.

 \subsection*{Limitation to the number of interleaved interferometers}
When trying 5 interleaved cycles, we observed a dramatic loss of contrast of the interferometer. The reason is that when a (descending) atom cloud at the output of the interferometer enters the detection region, a part of the light  scattered by the atoms   is directed towards the (ascending) cloud, which optically pumps atoms to unwanted magnetic states and heats them before they enter the interferometer.

\section*{Supplementary Materials}
Supplementary material for this article is available and contains:\\
 Supplementary text, Sections S1 to S6\\
 Figures S1 to S5\\
 Table S1\\
References \cite{Lautier2014,Cheinet2008,Stockton2011}

% **************************************

\label{biblio}
\bibliographystyle{ScienceAdvances}

\begin{thebibliography}{10}

\bibitem{Sheng2013}
D.~Sheng, S.~Li, N.~Dural, M.~V. Romalis, Subfemtotesla scalar atomic
  magnetometry using multipass cells.
\newblock {\it Phys. Rev. Lett.\/} {\bf 110}, 160802 (2013).

\bibitem{Gross2017}
I.~Gross, W.~Akhtar, V.~Garcia, L.~J. Martínez, S.~Chouaieb, K.~Garcia,
  C.~Carrétéro, A.~Barthélémy, P.~Appel, P.~Maletinsky, J.-V. Kim, J.~Y.
  Chauleau, N.~Jaouen, M.~Viret, M.~Bibes, S.~Fusil, V.~Jacques, Real-space
  imaging of non-collinear antiferromagnetic order with a single-spin
  magnetometer.
\newblock {\it Nature\/} {\bf 549}, 252 (2017).

\bibitem{Jimenez-Martinez2018}
R.~Jim\'enez-Mart\'{\i}nez, J.~Ko\l{}ody\ifmmode~\acute{n}\else \'{n}\fi{}ski,
  C.~Troullinou, V.~G. Lucivero, J.~Kong, M.~W. Mitchell, Signal tracking
  beyond the time resolution of an atomic sensor by kalman filtering.
\newblock {\it Phys. Rev. Lett.\/} {\bf 120}, 040503 (2018).

\bibitem{Takamoto2005}
M.~Takamoto, F.-L. Hong, R.~Higashi, H.~Katori, An optical lattice clock.
\newblock {\it Nature\/} {\bf 435}, 321 (2005).

\bibitem{LeTargat2013}
R.~Le~Targat, L.~Lorini, Y.~Le~Coq, M.~Zawada, J.~Guéna, M.~Abgrall, M.~Gurov,
  P.~Rosenbusch, D.~G. Rovera, B.~Nagórny, R.~Gartman, P.~G. Westergaard,
  M.~E. Tobar, M.~Lours, G.~Santarelli, A.~Clairon, S.~Bize, P.~Laurent,
  P.~Lemonde, J.~Lodewyck, Experimental realization of an optical second with
  strontium lattice clocks.
\newblock {\it Nature Communications\/} {\bf 4}, 2109 (2013).

\bibitem{Tarallo2014}
M.~G. Tarallo, T.~Mazzoni, N.~Poli, D.~V. Sutyrin, X.~Zhang, G.~M. Tino, Test
  of einstein equivalence principle for 0-spin and half-integer-spin atoms:
  Search for spin-gravity coupling effects.
\newblock {\it Phys. Rev. Lett.\/} {\bf 113}, 023005 (2014).

\bibitem{Bagci2014}
T.~Bagci, A.~Simonsen, S.~Schmid, L.~G. Villanueva, E.~Zeuthen, J.~Appel, J.~M.
  Taylor, A.~S{\o}rensen, K.~Usami, A.~Schliesser, E.~S. Polzik, Optical
  detection of radio waves through a nanomechanical transducer.
\newblock {\it Nature\/} {\bf 507}, 81 (2014).

\bibitem{Facon2016}
A.~Facon, E.-K. Dietsche, D.~Grosso, S.~Haroche, J.-M. Raimond, M.~Brune,
  S.~Gleyzes, A sensitive electrometer based on a rydberg atom in a
  schrödinger-cat state.
\newblock {\it Nature\/} {\bf 535}, 262 (2016).

\bibitem{Borde1989}
C.~Bord\'e, Atomic interferometry with internal state labelling.
\newblock {\it Physics Letters A\/} {\bf 140}, 10--12 (1989).

\bibitem{KasevichChu1991}
M.~Kasevich, S.~Chu, Atomic interferometry using stimulated raman transitions.
\newblock {\it Phys. Rev. Lett.\/} {\bf 67}, 181--184 (1991).

\bibitem{Riehle1991}
F.~Riehle, T.~Kisters, A.~Witte, J.~Helmcke, C.~J. Bord\'e, Optical ramsey
  spectroscopy in a rotating frame: Sagnac effect in a matter-wave
  interferometer.
\newblock {\it Phys. Rev. Lett.\/} {\bf 67}, 177--180 (1991).

\bibitem{Bouchendira2011}
R.~Bouchendira, P.~Clad\'e, S.~Guellati-Khélifa, F.~Nez, F.~Biraben, New
  {Determination} of the {Fine} {Structure} {Constant} and {Test} of the
  {Quantum} {Electrodynamics}.
\newblock {\it Physical Review Letters\/} {\bf 106} (2011).

\bibitem{Lepoutre2012}
S.~Lepoutre, A.~Gauguet, G.~Tr\'enec, M.~B\"uchner, J.~Vigu\'e,
  He-mckellar-wilkens topological phase in atom interferometry.
\newblock {\it Phys. Rev. Lett.\/} {\bf 109}, 120404 (2012).

\bibitem{Zhou2015}
L.~Zhou, S.~Long, B.~Tang, X.~Chen, F.~Gao, W.~Peng, W.~Duan, J.~Zhong,
  Z.~Xiong, J.~Wang, Y.~Zhang, M.~Zhan, Test of {Equivalence} {Principle} at
  $10^{-8}$ {Level} by a {Dual}-{Species} {Double}-{Diffraction} {Raman} {Atom}
  {Interferometer}.
\newblock {\it Physical Review Letters\/} {\bf 115} (2015).

\bibitem{Jaffe2017}
M.~Jaffe, P.~Haslinger, V.~Xu, P.~Hamilton, A.~Upadhye, B.~Elder, J.~Khoury,
  H.~Müller, Testing sub-gravitational forces on atoms from a miniature
  in-vacuum source mass.
\newblock {\it Nature Physics\/} {\bf 13}, 938--942 (2017).

\bibitem{Asenbaum2017}
P.~Asenbaum, C.~Overstreet, T.~Kovachy, D.~D. Brown, J.~M. Hogan, M.~A.
  Kasevich, Phase {Shift} in an {Atom} {Interferometer} due to {Spacetime}
  {Curvature} across its {Wave} {Function}.
\newblock {\it Physical Review Letters\/} {\bf 118} (2017).

\bibitem{Rosi2014}
G.~Rosi, F.~Sorrentino, L.~Cacciapuoti, M.~Prevedelli, G.~M. Tino, Precision
  measurement of the {Newtonian} gravitational constant using cold atoms.
\newblock {\it Nature\/} {\bf 510}, 518--521 (2014).

\bibitem{Peters2001}
A.~Peters, K.~Y. Chung, S.~Chu, High-precision gravity measurements using atom
  interferometry.
\newblock {\it Metrologia\/} {\bf 38}, 25 (2001).

\bibitem{Freier2016}
C.~Freier, M.~Hauth, V.~Schkolnik, B.~Leykauf, M.~Schilling, H.~Wziontek, H.-G.
  Scherneck, J.~M\"{u}ller, A.~Peters, Mobile quantum gravity sensor with
  unprecedented stability.
\newblock {\it J. Phys. Conf. Ser.\/} {\bf 723}, 012050 (2016).

\bibitem{Bidel2018}
Y.~Bidel, N.~Zahzam, C.~Blanchard, A.~Bonnin, M.~Cadoret, A.~Bresson,
  D.~Rouxel, M.~F. Lequentrec-Lalancette, Absolute marine gravimetry with
  matter-wave interferometry.
\newblock {\it Nature Communications\/} {\bf 9}, 627 (2018).

\bibitem{Karcher2018}
R.~{Karcher}, A.~{Imanaliev}, S.~{Merlet}, F.~{Pereira dos Santos}, {Improving
  the accuracy of atom interferometers with ultracold sources}.
\newblock {\it ArXiv e-prints\/}  (2018).

\bibitem{Lautier2014}
J.~Lautier, L.~Volodimer, T.~Hardin, S.~Merlet, M.~Lours, F.~Pereira
  Dos~Santos, A.~Landragin, Hybridizing matter-wave and classical
  accelerometers.
\newblock {\it Applied Physics Letters\/} {\bf 105}, 144102 (2014).

\bibitem{Jekeli2005}
C.~Jekeli, Navigation {Error} {Analysis} of {Atom} {Interferometer} {Inertial}
  {Sensor}.
\newblock {\it Navigation\/} {\bf 52}, 1--14 (2005).

\bibitem{Graham2016}
P.~W. Graham, D.~E. Kaplan, J.~Mardon, S.~Rajendran, W.~A. Terrano, Dark matter
  direct detection with accelerometers.
\newblock {\it Phys. Rev. D\/} {\bf 93}, 075029 (2016).

\bibitem{Chaibi2016}
W.~Chaibi, R.~Geiger, B.~Canuel, A.~Bertoldi, A.~Landragin, P.~Bouyer, Low
  frequency gravitational wave detection with ground-based atom interferometer
  arrays.
\newblock {\it Physical Review D\/} {\bf 93} (2016).

\bibitem{Graham2016a}
P.~W. Graham, J.~M. Hogan, M.~A. Kasevich, S.~Rajendran, Resonant mode for
  gravitational wave detectors based on atom interferometry.
\newblock {\it Phys. Rev. D\/} {\bf 94}, 104022 (2016).

\bibitem{Gauguet2009}
A.~Gauguet, B.~Canuel, T.~L\'ev\`eque, W.~Chaibi, A.~Landragin,
  Characterization and limits of a cold-atom sagnac interferometer.
\newblock {\it Phys. Rev. A\/} {\bf 80}, 063604 (2009).

\bibitem{Berg2015}
P.~Berg, S.~Abend, G.~Tackmann, C.~Schubert, E.~Giese, W.~Schleich,
  F.~Narducci, W.~Ertmer, E.~Rasel, Composite-{Light}-{Pulse} {Technique} for
  {High}-{Precision} {Atom} {Interferometry}.
\newblock {\it Physical Review Letters\/} {\bf 114} (2015).

\bibitem{Gustavson2000}
T.~L. Gustavson, A.~Landragin, M.~A. Kasevich, Rotation sensing with a dual
  atom-interferometer sagnac gyroscope.
\newblock {\it Classical and Quantum Gravity\/} {\bf 17}, 2385 (2000).

\bibitem{Durfee2006}
D.~S. Durfee, Y.~K. Shaham, M.~A. Kasevich, Long-{Term} {Stability} of an
  {Area}-{Reversible} {Atom}-{Interferometer} {Sagnac} {Gyroscope}.
\newblock {\it Physical Review Letters\/} {\bf 97} (2006).

\bibitem{Lefevre2014}
H.~C. Lef\`evre, The fiber-optic gyroscope, a century after {Sagnac}'s
  experiment: {The} ultimate rotation-sensing technology?
\newblock {\it Comptes Rendus Physique\/} {\bf 15}, 851--858 (2014). ; For
  recent performances, see, e.g.
  \href{http://web.ixblue.com/cn/aw6ym/fiberoptic-gyroscope}{iXblue
  ultimate-performance Fiber-Optic Gyroscope (FOG)}.

\bibitem{Dutta2016}
I.~Dutta, D.~Savoie, B.~Fang, B.~Venon, C.~Garrido~Alzar, R.~Geiger,
  A.~Landragin, Continuous {Cold}-{Atom} {Inertial} {Sensor} with 1 nrad / sec
  {Rotation} {Stability}.
\newblock {\it Physical Review Letters\/} {\bf 116} (2016).

\bibitem{Canuel2006}
B.~Canuel, F.~Leduc, D.~Holleville, A.~Gauguet, J.~Fils, A.~Virdis, A.~Clairon,
  N.~Dimarcq, C.~J. Bord\'e, A.~Landragin, P.~Bouyer, Six-axis inertial sensor
  using cold-atom interferometry.
\newblock {\it Phys. Rev. Lett.\/} {\bf 97}, 010402 (2006).

\bibitem{Stockton2011}
J.~K. Stockton, K.~Takase, M.~A. Kasevich, Absolute {Geodetic} {Rotation}
  {Measurement} {Using} {Atom} {Interferometry}.
\newblock {\it Physical Review Letters\/} {\bf 107} (2011).

\bibitem{Legere1998}
R.~Legere, K.~Gibble, Quantum scattering in a juggling atomic fountain.
\newblock {\it Phys. Rev. Lett.\/} {\bf 81}, 5780--5783 (1998).

\bibitem{Meunier2014}
M.~Meunier, I.~Dutta, R.~Geiger, C.~Guerlin, C.~L. Garrido~Alzar, A.~Landragin,
  Stability enhancement by joint phase measurements in a single cold atomic
  fountain.
\newblock {\it Physical Review A\/} {\bf 90} (2014).

\bibitem{Dickerson2013}
S.~M. Dickerson, J.~M. Hogan, A.~Sugarbaker, D.~M.~S. Johnson, M.~A. Kasevich,
  Multiaxis {Inertial} {Sensing} with {Long}-{Time} {Point} {Source} {Atom}
  {Interferometry}.
\newblock {\it Physical Review Letters\/} {\bf 111} (2013).

\bibitem{Hensley1999}
J.~M. Hensley, A.~Peters, S.~Chu, Active low frequency vertical vibration
  isolation.
\newblock {\it Review of Scientific Instruments\/} {\bf 70}, 2735--2741 (1999).

\bibitem{Tackmann2012}
G.~Tackmann, P.~Berg, C.~Schubert, S.~Abend, M.~Gilowski, W.~Ertmer, E.~M.
  Rasel, Self-alignment of a compact large-area atomic sagnac interferometer.
\newblock {\it New Journal of Physics\/} {\bf 14}, 015002 (2012).

\bibitem{Hu2017}
L.~Hu, N.~Poli, L.~Salvi, G.~M. Tino, Atom interferometry with the sr optical
  clock transition.
\newblock {\it Phys. Rev. Lett.\/} {\bf 119}, 263601 (2017).

\bibitem{Rudolph2015}
J.~Rudolph, W.~Herr, C.~Grzeschik, T.~Sternke, A.~Grote, M.~Popp, D.~Becker,
  H.~Müntinga, H.~Ahlers, A.~Peters, C.~Lämmerzahl, K.~Sengstock, N.~Gaaloul,
  W.~Ertmer, E.~M. Rasel, A high-flux bec source for mobile atom
  interferometers.
\newblock {\it New Journal of Physics\/} {\bf 17}, 065001 (2015).

\bibitem{Cheinet2008}
P.~Cheinet, B.~Canuel, F.~P.~D. Santos, A.~Gauguet, F.~Yver-Leduc,
  A.~Landragin, Measurement of the sensitivity function in a time-domain atomic
  interferometer.
\newblock {\it IEEE Transactions on Instrumentation and Measurement\/} {\bf
  57}, 1141--1148 (2008).

\end{thebibliography}

%%% ****************************

\section*{Acknowledgments}
\label{Acknowledgments}
We thank F. Pereira Dos Santos for a careful reading of the manuscript. 

Funding: We acknowledge the financial support from Ville de Paris (project HSENS-MWGRAV), FIRST-TF (ANR-10-LABX-48-01),   Centre National d'Etudes Saptiales (CNES), Sorbonne Universit\'es (project SU-16-R-EMR-30, LORINVACC) and Action Sp\'ecifique du CNRS Gravitation, R\'ef\'erences, Astronomie et M\'etrologie (GRAM). B.F. was funded by Conseil
Scientifique de l'Observatoire de Paris, D.S. by Direction G\'en\'erale de l'Armement, and M.A. by the EDPIF doctoral school. 

Author contributions: D.S., M.A., B.F. performed the experiments and L.A.S. contributed to the dynamic rotation rate measurements.
D.S., R.G. and M.A. analyzed the data. R.G. and D.S. wrote the manuscript.
A.L. conceived the experiment.
R.G. and A.L. supervised the research.
All authors discussed the manuscript.

Competing Interests: The authors declare that they have no competing interests.  

Data and materials availability: All data needed to evaluate the conclusions in the paper are present in the paper and the Supplementary Materials.  Additional data available from authors upon request.

\newpage

\begin{center}
\Large{\textbf{Supplementary Materials}}
\end{center}

\subsection*{Section S1: Real-time compensation of vibration noise, mid-fringe lock, and details of the sequence } 
\label{sec:RTC}
We calculate in real time the linear acceleration phase of the atom interferometer and compensate for it by applying a phase jump $\Phi_{\text{RTC}}$ at each cycle in the phase lock loop (PLL) of the Raman lasers \cite{Lautier2014}. Specifically, we record, for each interferometer cycle, the signals of two seismometers during $786$~ms, and compute the vibration phase shift by applying the sensitivity function of the 4 pulse interferometer to the half sum of the seismometer signals. 
As the transfer of the phase jump information to the direct digital synthesizer driving the PLL takes about 10~ms, we stop the seismometer acquisition 15~ms before the end of the cycle (yielding $801-15=786$~ms). This corresponds to an (estimated) 60~mrad rms error in the evaluation of the vibration phase. 

\begin{figure}[!h]
\centering
\includegraphics[width=0.9\linewidth]{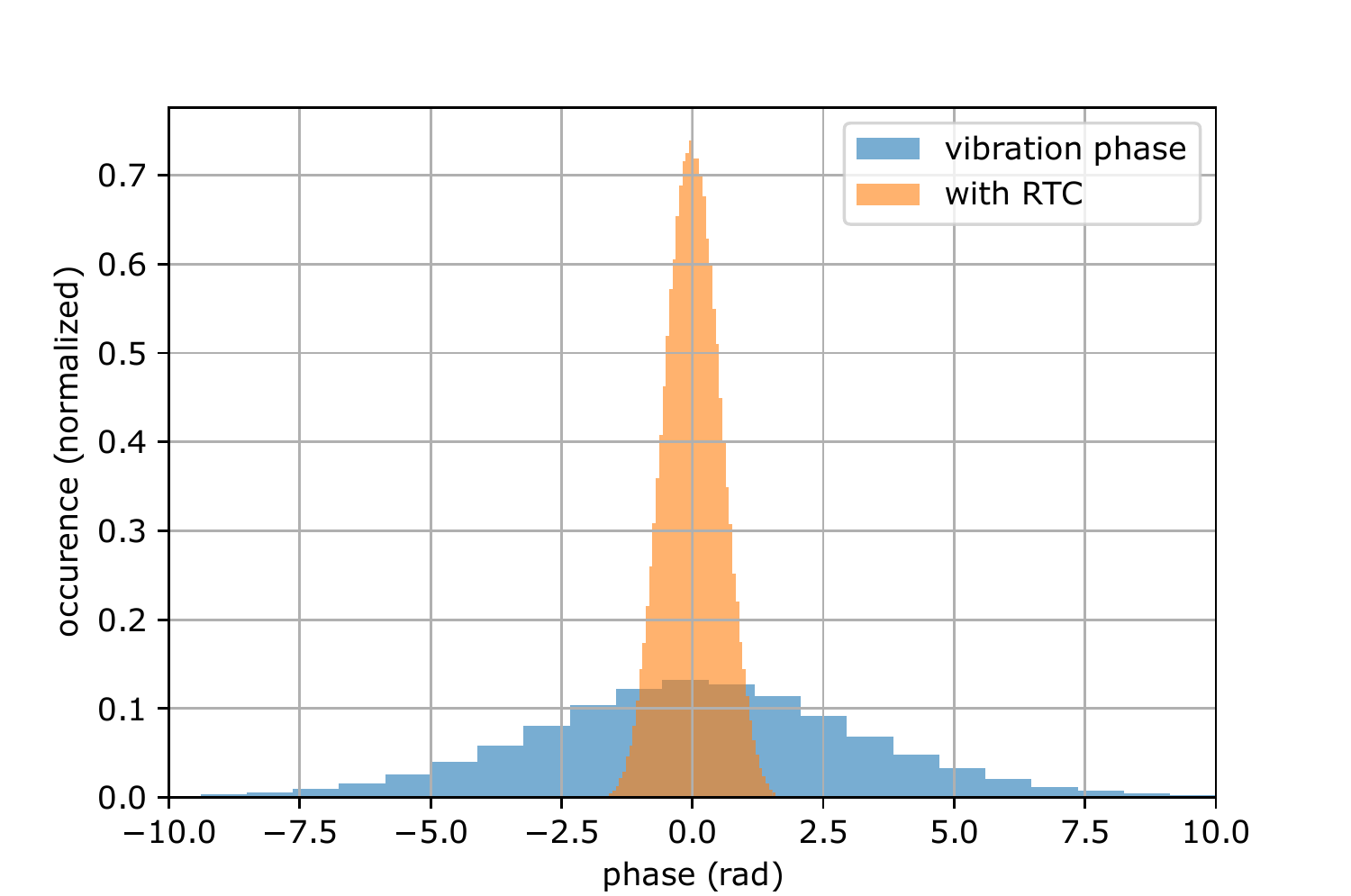}
\caption{\label{fig:histPhaseRTC} 
Histogram of the vibration phase  and of the interferometer phase with real time compensation (RTC) of vibration. The standard deviation of the vibration phase (interferometer phase) is 3.2 (0.52)~rad.
}
\end{figure}

Fig.~\ref{fig:histPhaseRTC} shows the histogram (in blue) of the estimated vibration phase, corresponding to a phase noise of 3.2~rad in standard deviation. The real time compensation of vibration allows to work within the linear response of the interferometer by reducing the phase spread to 0.52~rad, i.e. by a factor of 6 (orange histogram).

\begin{figure}[!h]
\centering
\includegraphics[width=0.9\linewidth]{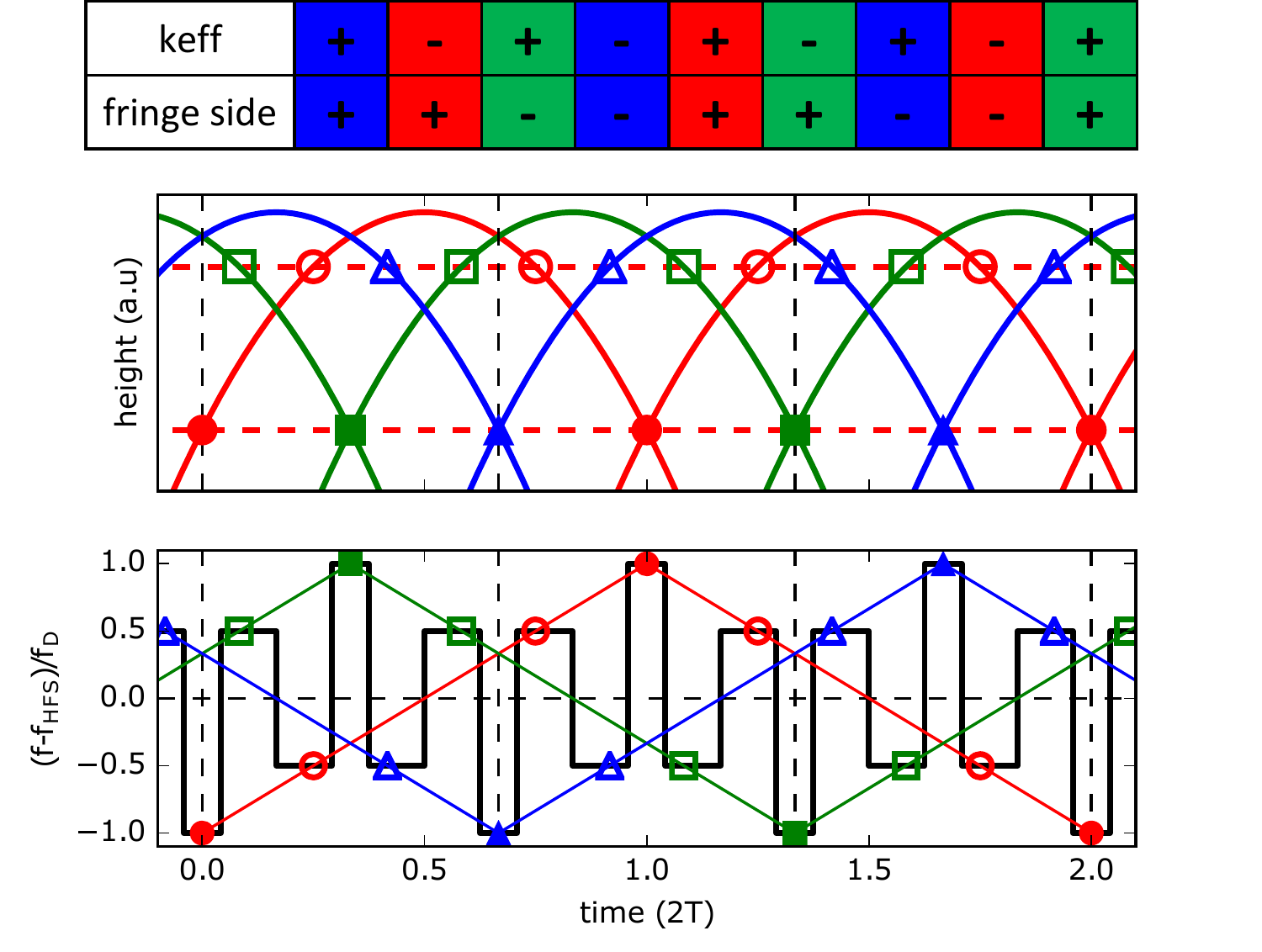}
\caption{\label{fig:sequence_keff_fringeSide} 
\textbf{Details of the sequence.} See the text for the explanations of the panels. In the bottom panel, $(f-f_{\text{HFS}})$ denotes the  offset of the Raman laser relative frequencies from the hyperfine splitting frequency, and $f_\text{D}\simeq 611$~kHz the maximum Doppler effect, as defined in the Methods of the main text.
%Top panel: Detail of the sequence showing the alternation of $\pm \veckeff$ and of the fringe side. Middle panel:  trajectories (height versus time) of the successively interrogated atom clouds (parabolic lines) and light pulses (symbols). Bottom panel:  relative frequency of the Raman lasers over time (thick black line) to fulfill the two-photon resonance condition (thin colored lines), with the light pulses  marked by the symbols.  
}
\end{figure}

 To work around mid-fringe on long-term acquisitions, an extra phase jump $\Phi_{\mathrm{SRS},i}=\Phi_{\mathrm{MFL},i} + (-1)^i \times \frac{\pi}{2}$ is added to $\Phi_{\text{RTC},i}$ at each cycle $i$. The value of the mid-fringe lock (MFL) phase is  servoed  by alternating measurements on both sides of a fringe with the application of successive $\pm \pi/2$ phase shifts (see Fig.~\ref{fig:sequence_keff_fringeSide}, top panel). 
 At the $i^\mathrm{th}$ cycle, the error signal of the loop is calculated as
\begin{equation}
\epsilon_i = (-1)^i(P_i - P_{i-1}),
\end{equation}
where $P_i$ is the transition probability at the $i^\mathrm{th}$ cycle. The correction is a pure integrator, with the value of the phase at cycle  $i$ given by:
\begin{equation}
\Phi_{\mathrm{MFL},i} = G\sum_{p=0}^i \epsilon_p
\end{equation}
where $G$ is the gain of the lock. The value $\Phi_{\mathrm{MFL},i}$ is calculated at cycle $i$ and applied at cycle $i+1$.
Once the lock has converged (typically after  50~s for a gain $G=30$), the value of the interferometer phase can be directly read from $\Phi_\mathrm{MFL}$. The total phase of the atom interferometer (for each value of $\veckeff$) is computed as follows:
\begin{equation}
\Phi_i =  (-1)^i\frac{P_i - P_{i-1}}{2A} + \frac{1}{2}\left( \Phi_{\mathrm{SRS},i} + \Phi_{\mathrm{SRS},i-1}\right),
\label{eq:total_phase}
\end{equation}
where $A$ is the fringe amplitude.
  The stability of the interferometer phase (as analyzed in Fig.~3 of the main text) is mainly given by the first term of Eq.~\eqref{eq:total_phase} at short term (for integration times below the mid-fringe locking point), while the second term is representative of the phase stability once the lock has converged.
  
Fig.~\ref{fig:sequence_keff_fringeSide} shows the detail of the sequence.  The top panel illustrates the alternation of the Raman wavevector ($\pm \veckeff$) and of the interferometer fringe side.  The middle panel shows the height of the atomic cloud as a function of time (solid lines), thereby indicating the shared $\pi/2$ light pulses (solid symbols) and the not shared $\pi$ light pulses (empty symbols).  The color code highlights the joint points : successive sequences of the same color have one $\pi/2$ pulse in common.  The bottom panel plots the corresponding frequency difference of the Raman lasers (thick black line) to meet the two-photon resonance condition (thin colored lines, equivalent to a linear frequency ramp) at each shot. The symbols mark the light pulses.

\newpage

\subsection*{Section S2: Raw data}
\begin{figure}[!h]
\centering
\includegraphics[width=0.95\linewidth]{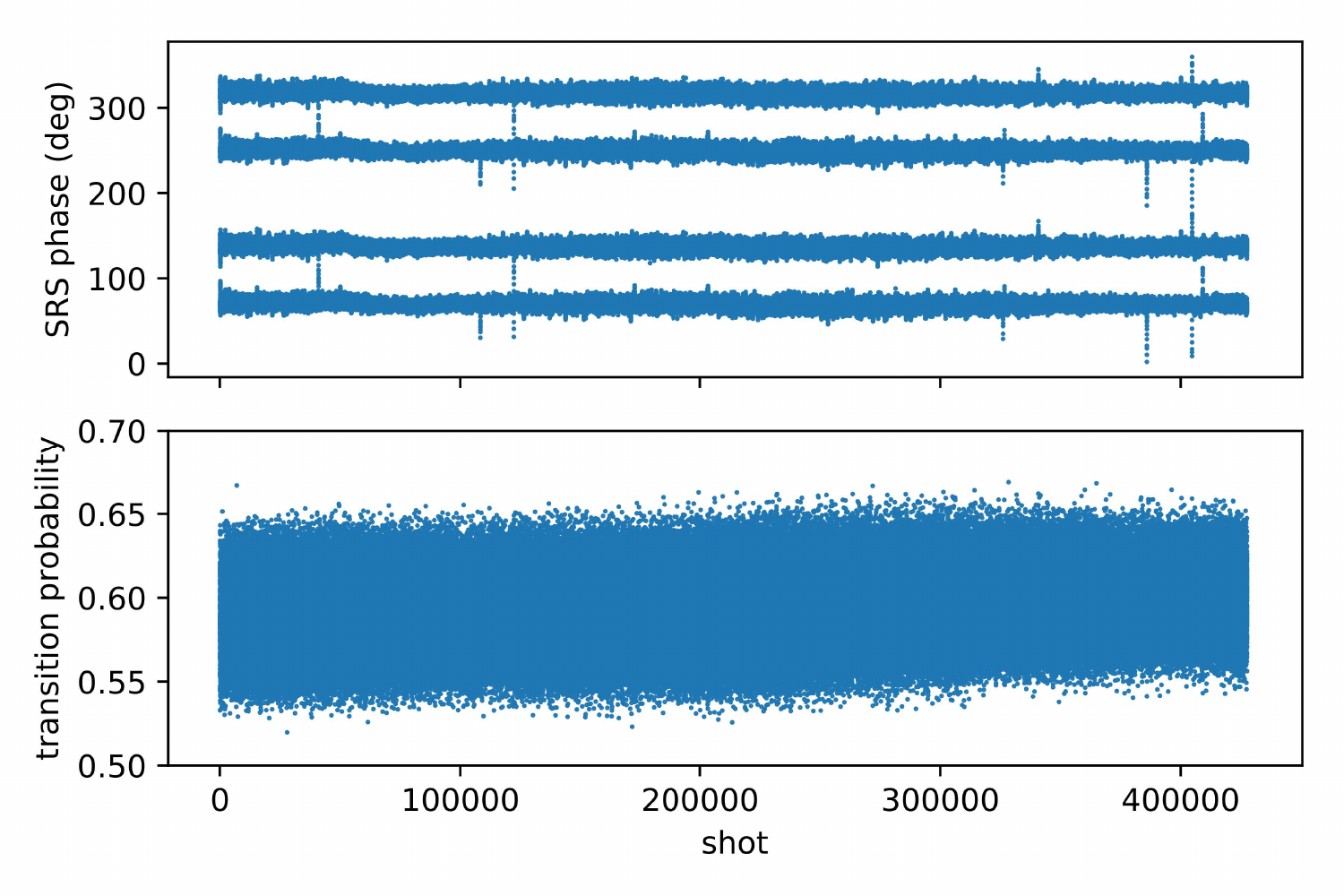}
\caption{\textbf{Raw interferometer measurements corresponding to the data presented in Fig.~2 of the main text.} One shot corresponds to $T_c=0.267$~s. The time axis spans from 21:10 on 23/09/2018 to 5:41 on 25/09/2018.
% data set 13 =  21:10 on 23/09 to 5:41 on 25/09
The top panel shows the phase of the direct digital synthetizer (SRS) that is used to steer the interferometer at mid-fringe, for the four different configurations of interferometer ($\pm \veckeff$ and each with two sides of the fringe). The bottom panel shows the atomic transition probability, $P$, from which we extract the  phase following Eq.~\eqref{eq:total_phase} with  $A\simeq 0.037$ the fringe amplitude. 
%The total atom interferometer phase is the sum of the SRS phase and of the phase obtained from the transition probability. 
}
\label{figS:raw_data}
\end{figure}

Fig.~\ref{figS:raw_data} shows the raw data from which we extract the inertial phase shown in the main text. 
As written in Eq.~\eqref{eq:total_phase}, the total phase is computed as a sum of the SRS phase and of the atomic phase obtained from the transition probability. 
We extract the atomic phase by using a linear approximation of the response of the atom interferometer instead of the sinusoidal response, since  we operate near the center of the fringe. 
%We have verified that the linear approximation gives a good estimation of the phase, with a fidelity of $95\%$.
%Fig.~\ref{figS:phaseLinearVsArcsin} shows that the linear approximation gives a good estimation of the phase.

%\begin{figure}[!h]
%\centering
%\includegraphics[width=0.8\linewidth]{Figures_SM/13_hist_phase_proba_3_sigma.pdf}
%\caption{\label{figS:phaseLinearVsArcsin} \textcolor{red}{Should we show this figure ?} Extracting the phase from the transition probability using a linear approximation or using arcsin.}
%\end{figure}

\newpage

\subsection*{Section S3: Analysis of vibration noise}
\begin{figure}[!h]
\centering
\includegraphics[width=\linewidth]{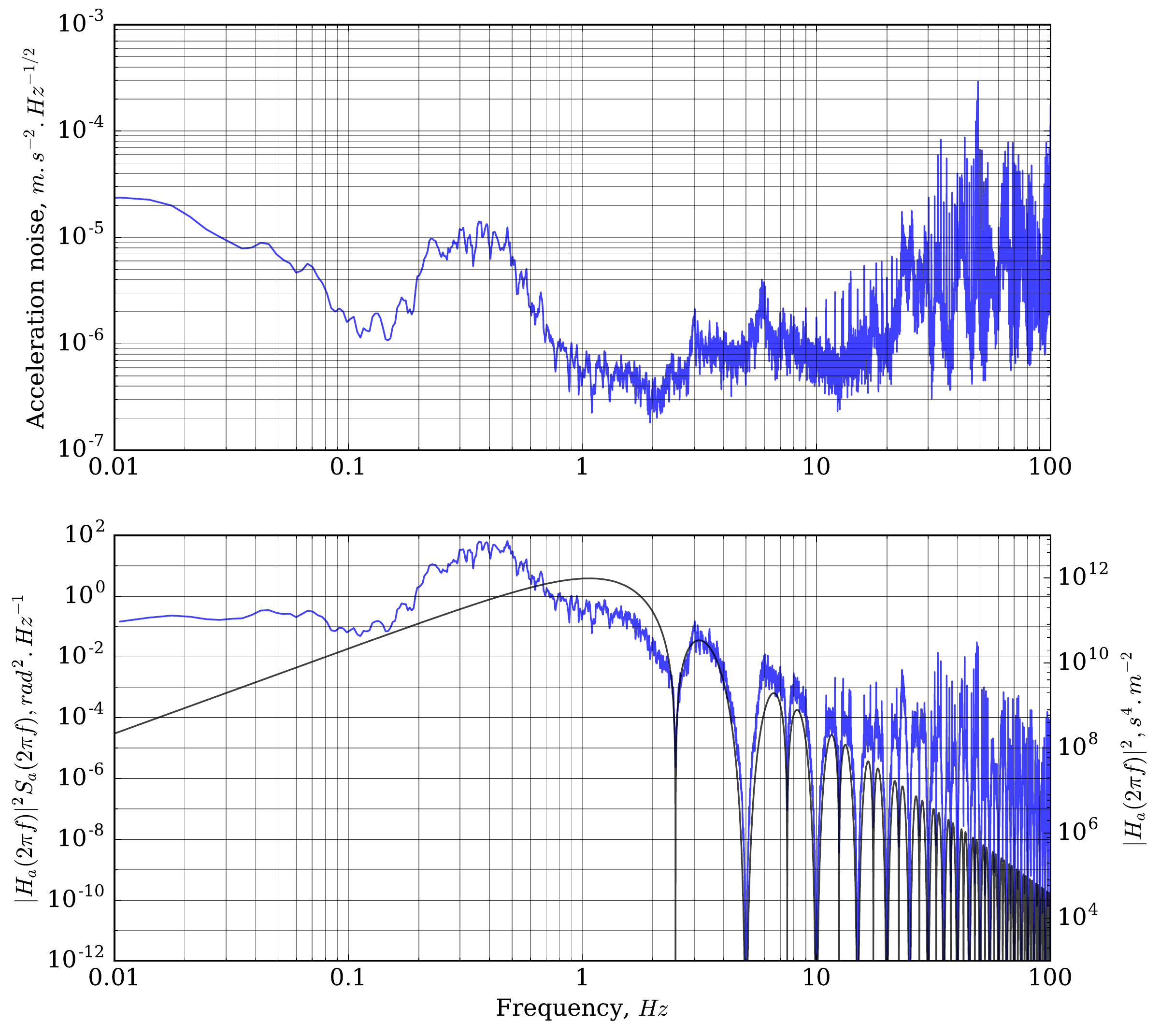}
\caption{\label{figS:vibNoiseSpect}
\textbf{Analysis of vibration noise.} Top: Amplitude spectral density of linear acceleration noise. Bottom: contribution to interferometer phase noise as from Eq.~\eqref{eq:sigmaPhi2}. The black line shows the acceleration transfer function of the interferometer, $|H_a(2\pi f)|^2$. }
\end{figure}

Fig.~\ref{figS:vibNoiseSpect} shows the amplitude spectral density of the linear acceleration noise of the experiment,$\sqrt{S_a(2\pi f)}$, recorded by the two seismometers. We estimate the contribution of each frequency band $[f_1-f_2]$ to the interferometer phase noise  by 
\begin{equation}
\sigma_\Phi^2(f_1,f_2) = \int_{f_1}^{f_2} |H_a(2\pi f)|^2 S_a(2\pi f) df,
\label{eq:sigmaPhi2}
\end{equation} 
with $|H_a(2\pi f)|= \frac{8 k_{\text{eff}}}{\omega^2 }  \sin(\frac{\omega T}{2}) \sin^2(\frac{\omega T}{4})$ the transfer function of the 4 pulse interferometer, see Ref.~\cite{Cheinet2008}. This transfer function acts as a band-pass filter with a peak sensitivity centered around $1/T$. 
The result of the numerical integration in Eq.~\eqref{eq:sigmaPhi2} is given in Table~\ref{TabS:sigma_freq_band}.
The main contribution to the interferometer phase noise is  the frequency band centered around 0.5~Hz.
The estimation of the total phase noise  (3.2~rad) with this method matches with the measured standard deviation of vibration phase (Fig.~\ref{fig:histPhaseRTC}).

\begin{table}[!h]
\centering
\begin{tabular}{|c|c|c|c|c|c|c|c|c|}
\hline
frequency band (Hz) & 0.01-0.1 & 0.1-0.3 & 0.3-0.4 & 0.4-0.5 &0.5-1 & 1-10 & 10-100 & total\\
\hline
$\sigma_\Phi^2$ (rad$^2$) & 0.02 & 0.93 & 3.7 & 3.9 & 1.4 & 0.26 & 0.005 & 10.4\\
\hline
\end{tabular}
\caption{Contribution of the linear acceleration noise to the interferometer phase noise by frequency band. The total rms phase noise is $\sigma_\Phi=3.2$~rad.}
\label{TabS:sigma_freq_band}
\end{table}

\newpage

\newpage

\subsection*{Section S4: Stability analysis}
\begin{figure}[!h]
\centering
\includegraphics[width=\linewidth]{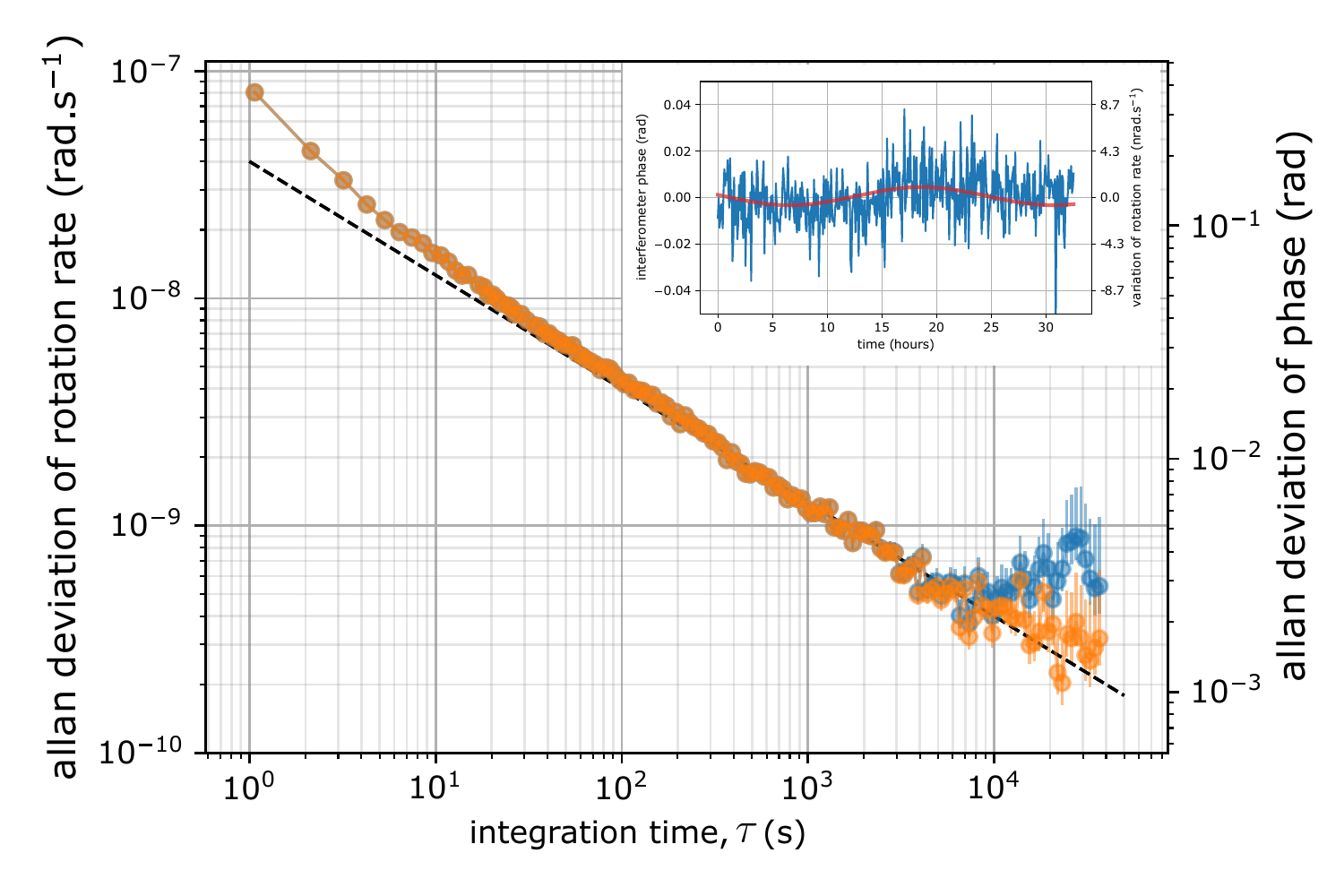}
\caption{\label{figS:adevLT} \textbf{Stability analysis of the gyroscope}. 
Blue: Allan deviation of the rotation rate measurement  of Fig.~2 of the main text. 
Orange: Allan deviation of the measurement when subtracting a fitted sinusoidal modulation of fixed 24 hour period (inset).   The dashed line is a visual guide to the eye representing a scaling as $4\times 10^{-8}  \ \text{rad.s}^{-1} \times \tau^{-1/2} $. 
}
\end{figure}

%After this systematic analysis, we recorded the rotation rate acquisition displayed on Fig.~\ref{fig:long_acq}. 
The  stability of the gyroscope over the entire acquisition displayed in Fig.~2 of the main text is  analyzed in Fig.~\ref{figS:adevLT}.
As the data set includes the day time were the vibration noise is higher than during night time, the short term sensitivity is slightly worse than in Fig.~3 of the main text, but the $\tau^{-1}$ scaling is still clearly visible.

When looking closely at the raw data averaged over 267~s segments (orange trace in Fig.~2 of the main text), a small modulation can be noticed. To highlight it, we show in the inset of Fig.~\ref{figS:adevLT} the result of a sinusoidal fit to the 267~s averaged data  where the fit period is fixed to 24 hours; the fitted amplitude equals 4.0~mrad.
This sinusoidal modulation reveals itself as a drift of the Allan deviation around 10~000 seconds, at the level of about 0.5~nrad.s$^{-1}$.
We attribute this modulation to a daily variation  of the relative angle between the Raman beams in the $y$ direction: as explained in the Material and Methods (section 'Alignment of the two Raman beams and atom trajectory'), a drift of $0.2$~$\mu$rad is  enough to explain a modulation of $4.0$~mrad of the interferometer phase.
Removing this instrumental drift yields the orange trace in Fig.~\ref{figS:adevLT}, showing  a stability of \LongTermStability \ at 30~000~s integration time, in agreement with that read from Fig.~3 of the main text for a shorter integration time.

\subsection*{Section S5: Analysis of the dynamic rotation rate measurements}
Following Eq.~1 of the main text and Ref.~\cite{Stockton2011}, the full phase shift  in our atom interferometer can be expressed as
\begin{equation}
\Phi = \frac{1}{2}\veckeff\cdot (\vec{\Omega}_E\times \vec{g})T^3 +  \frac{3}{4}\veckeff\cdot (\vec{\Omega}_F\times \vec{g} + \vec{\Omega}_E\times \vec{a}  + \vec{\Omega}_F\times \vec{a})T^3,
\label{eq:dynamic_meas}
\end{equation}
with $\vec{\Omega}_E$ the Earth rotation rate, $\vec{g}$ the local acceleration of gravity, $\vec{\Omega}_F$ the rotation rate of the experimental apparatus with respect to the local geostationary reference frame, and $\vec{a}$ the acceleration of the apparatus with respect to that frame.
For the demonstration of dynamic rotation rate measurements presented in the main text (Fig.~4), the rotation rate of the apparatus is modulated as $\vec{\Omega}_F(t)=\Omega_0\cos(\omega t)\hat{u}_y$, the Earth rotation phase shift (first term in Eq.~\eqref{eq:dynamic_meas}) is a constant offset, and the two terms in the second parenthesis involving the acceleration $\vec{a}$ of the apparatus are negligible. Therefore the dynamical phase shift reads
\begin{equation}
\Phi_{\text{dyn}}(t) \simeq  \frac{3}{4}\veckeff\cdot ( \vec{\Omega}_F(t)\times \vec{g} )T^3=\mathcal{S}\Omega_F(t),
\end{equation}
with $\mathcal{S}=\frac{3}{4}\keff g T^3$.
This expression is valid if the rotation rate is constant over the duration of the interferometer ($2T=801$~ms).
Here, we apply  sinusoidal modulations of $2\pi/\omega=5-10$~s period, and the atom interferometer delivers at every shot $i T_c$  a  rotation signal integrated over a duration of $2T$:
\begin{equation}
\Phi(iT_c)=\mathcal{S}\Omega_0\cos[\omega(iT_c+T)]\times \frac{\sin(\omega T)}{\omega T}.
\label{eq:expected_signal}
\end{equation}
The last term corresponds to a correction of $0.96$ for $5$~s period, and 0.99 for 10~s period.
To reconstruct the total phase from the atom interferometer, we sum the contribution of the phase deduced from the transition probability, $P_i/A$ (with $A$ the fringe amplitude), and the phase $\Phi_{\text{RTC},i}$ which was used in the real time compensation. 
%Fig.~4 a) and b) of the main text show the atomic phase, and Fig.4 c) shows the spectrum of the total phase. 
%\textcolor{red}{Say how was done the FFT (how many points).}

To compare the amplitude of the signal measured on the atom interferometer with the expected signal  amplitude, we need to know accurately the applied rotation rate amplitude $\Omega_0=\omega\theta_0$. Here $\theta_0$ is the amplitude of the applied sinusoidal modulation of the orientation of the experiment, as defined in the main text. We measure $\theta_0$ by comparing the signal of three different sensors: a tiltmeter, an accelerometer (which is sensitive to the the projection of  gravity on its horizontal axis, $g\sin\theta_x(t)\simeq g \theta_x(t)=g\theta_0\sin(\omega t)$), and a seismometer (which is sensitive to the integrated projection of gravity). Taking into account the calibrated response of these instruments, we extract $\theta_0$ and find an agreement between the three sensors with a relative discrepancy of $5\%$. This allows us to calculate the amplitude of the sinusoidal signal in Eq.~\eqref{eq:expected_signal},  $\mathcal{S}\omega\theta_0\times \frac{\sin(\omega T)}{\omega T}$, with an uncertainty of $5\%$.
 The  measured amplitude agree with the expected amplitude within this uncertainty  on  $\theta_0$.

\subsection*{Section S6: Systematic effect from the scattered light}
\label{sec:scatt_light}

The  light scattered by the MOT atoms, $ P_{\text{sc}}(t)$,  induces a differential light shift, $\delta_{\text{AC}}(t)\propto P_{\text{sc}}(t)$, which, when integrated over time along the trajectories of the atoms in the interferometer, results in a phase shift, $\phi_{\text{AC}}\propto \int  P_{\text{sc}}(t) dt$. 
The induced light shift is largely canceled by the symmetric, spin-echo-like, four pulse configuration, and by the use of $\veckeff$ reversal.
Still, to measure any possible residual effect from the scattered light on the inertial signal, we vary the MOT loading time from $t_1=35$~ms to $t_2=55$~ms. On such timescales, we observe that  $P_{\text{sc}}(t)\propto t$ by recording the MOT fluorescence, which corresponds to a scaling of the induced light shift $\phi_{\text{AC}}\propto t^2$. 

When varying the loading time from $t_1$ to $t_2$, we measure a non-inertial contribution (obtained from the half-sum of $\veckeff$ measurements) of $(55\pm 16)$~mrad, in agreement with a calculation of the expected light shift contribution.
On the inertial signal (obtained from the half-difference of $\veckeff$), the measured phase difference is $(20\pm 17)$~mrad.
%i.e. compatible with zero at the level of $4\times 10^{-9}$~\radps \ (we translate phase to rotation rate using the scale factor of Eq.~\eqref{eq:constant_rot}). 
Since $\phi_{\text{AC}}\propto t^2$, varying the loading time from $t_1$ to $t_2$ amounts to vary the systematic shift by a relative factor $[(t_2-t_1)/t_2]^2\simeq 60\%$, with respect to the effect associated to the usual $t_2=55$~ms loading time.
Our measured fluctuations of $1\%$ rms of the level of scattered light (for  55~ms MOT loading time) thus translate into  an upper bound on the phase instability of $20 \ \text{mrad}\times\frac{1}{60}\simeq0.3$~mrad, corresponding to a gyroscope instability of $7\times 10^{-11}$~\radps \ (we translate phase to rotation rate using the scale factor of Eq.~2 of the main text). The influence of the  scattered light from the fluorescence detection was evaluated to be even lower.

\end{document}